\documentclass[aps]{revtex4}   
\usepackage{amsmath}
\usepackage{amsfonts}
\usepackage{graphicx}         

\begin{document}


\newlength{\defbaselineskip}
\setlength{\defbaselineskip}{\baselineskip}
\newcommand{\setlinespacing}[1]
           {\setlength{\baselineskip}{#1 \defbaselineskip}}
\newcommand{\doublespacing}{\setlength{\baselineskip}
                           {2.0 \defbaselineskip}}
\newcommand{\singlespacing}{\setlength{\baselineskip}{\defbaselineskip}}

\newcommand{\eqn}[1]{Eq.~(\ref{#1})}
\newcommand{\Eqn}[1]{Equation~(\ref{#1})}
\newcommand{\fig}[1]{Fig.~\ref{#1}}
\newcommand{\Fig}[1]{Figure~\ref{#1}}
\newcommand{\secn}[1]{Section~\ref{#1}}
\newcommand{\Secn}[1]{Section~\ref{#1}}
\newcommand{\defn}[1]{definition~\ref{#1}}
\newcommand{\Defn}[1]{Definition~\ref{#1}}

\newcommand{\abs}[1]{|#1|}
\newcommand{\realpart}[1]{\,\mathrm{Re}\!\left\{#1\right\}}
\newcommand{\equivalent}{\Leftrightarrow}
\newcommand{\hence}{\Rightarrow}
\newcommand{\w}{\omega}
\newcommand{\scr}[1]{{\scriptscriptstyle{#1}}}
\newcommand{\sub}[1]{_{\mathrm{#1}}}
\newcommand{\supscr}[1]{^{\mathrm{#1}}}
\newcommand{\e}[1]{\hat{e}_{#1}}
\newcommand{\E}[1]{\times 10^{#1}}
\newcommand{\lsim}{\;$\raisebox{-0.5ex}{$\stackrel{<}{\scriptstyle{\sim}}$}$\;}
\newcommand{\gsim}{\;$\raisebox{-0.5ex}{$\stackrel{>}{\scriptstyle{\sim}}$}$\;}

\newcommand{\unitmatrix}{{\mathbf 1}}

\newcommand{\dnd}[3]{\frac{\partial^{#1} #2}{\partial #3^{#1}}}
\newcommand{\ddd}[2]{\dnd{2}{#1}{#2}}
\newcommand{\dd}[2]{\frac{\partial #1}{\partial #2}}
\newcommand{\ddt}[1]{\dd{#1}{t}}
\newcommand{\ddx}[1]{\dd{#1}{x}}
\newcommand{\ddy}[1]{\dd{#1}{y}}
\newcommand{\ddz}[1]{\dd{#1}{z}}

\newcommand{\V}[1]{\underline{#1}}
\newcommand{\grad}{\bigtriangledown}

\newcommand{\set}[1]{\{#1\}}
\newcommand{\R}{{\mathbb R}}
\newcommand{\C}{{\mathbb C}}

\newcommand{\units}[1]{\mathrm{\scriptstyle{#1}}}
\newcommand{\degrees}{^\circ}
\newcommand{\micron}{\mu\mathrm{m}}
\newcommand{\microns}{\mu\mathrm{m}}
\newcommand{\chem}[1]{$\mathrm{#1}$}

\newcommand{\state}[1]{|{\mathbf{#1}}\!\!>}
\newcommand{\dual}[1]{<\!\!{{\mathbf{#1}}}|}
\newcommand{\op}[1]{\mathbf{#1}}
\newcommand{\opdensity}[1]{{\mathbf{#1}}}
\newcommand{\ip}[2]{<\!\!{#1}|{#2}\!\!>}
\newcommand{\opip}[2]{\ip{\mathbf{#1}}{\mathbf{#2}}}
\newcommand{\T}{{}^{\dagger}}
\newcommand{\unitvec}[1]{\V{{\hat{#1}}}}

\newtheorem{example}{Example}

\title{Thermal noise in interferometric gravitational
wave detectors due to dielectric optical coatings}

\author{Gregory~M.~Harry\footnote{gharry@ligo.mit.edu}}
\address{LIGO Project, Massachusetts Institute of
Technology, Room~NW17-161, 175~Albany
Street, Cambridge, Massachusetts 02139, USA.\\
Department of Physics, Syracuse University, Syracuse, New York 13244-1130, USA.
\vspace{3ex}}

\author{Andri~M.~Gretarsson\footnote{andri@physics.syr.edu},
        Peter~R.~Saulson,
        Scott E. Kittelberger, Steven~D.~Penn,
        William~J.~Startin}
\address{Department of Physics, Syracuse University,
Syracuse, New York 13244-1130, USA. \vspace{3ex}}

\author{Sheila~Rowan, Martin~M.~Fejer}
\address{Edward L. Ginzton Lab,
Stanford University, Stanford, California 94305-4085, USA.
\vspace{3ex}}

\author{D. R. M. Crooks, Gianpietro~Cagnoli, Jim~Hough}
\address{Department of Physics and Astronomy,
University of Glasgow, Glasgow G12 8QQ, Scotland, United Kingdom.\vspace{2ex}}

\author{Norio Nakagawa}
\address{Center for Nondestructive Evaluation, Institute
for Physical Research and Technology, Iowa State University, Ames, Iowa 50011,
USA.\vspace{2ex}}

\date{\today}


\begin{abstract}
We report on thermal noise from the internal friction of dielectric coatings made from
alternating layers of Ta$_2$O$_5$ and SiO$_2$ deposited on fused silica substrates.  We
present calculations of the thermal noise in gravitational wave interferometers due to
optical coatings, when the material properties of the coating are different from those of
the substrate and the mechanical loss angle in the coating is anisotropic. The loss angle
in the coatings for strains parallel to the substrate surface was determined from
ringdown experiments. We measured the mechanical quality factor of three fused silica
samples with coatings deposited on them. The loss angle, $\phi_\|\left(f\right)$, of the
coating material for  strains parallel to the coated surface was found to be $4.2\pm
0.3\times 10^{-4}$ for coatings deposited on commercially polished slides and $1.0 \pm
0.3 \times 10^{-4}$ for a coating deposited on a superpolished disk. Using these numbers,
we estimate the effect of coatings on thermal noise in the initial LIGO and advanced LIGO
interferometers. We also find that the corresponding prediction for thermal noise in the
40~m LIGO prototype at Caltech is consistent with the noise data. These results are
complemented by results for a different type of coating, presented in a companion paper.
\end{abstract}

\maketitle
\section{Introduction}\label{Introduction}

The experimental effort to detect gravitational waves is entering an important phase.  A
number of interferometric gravitational wave observatories are being built around the
world~\cite{Abramovici,Virgo,GEO,TAMA} and most should be operational in the next few
years.  Plans are already being developed to operate the next generation of
interferometers; crucial research and development is going on now to ensure that these
interferometers will have the sensitivity necessary to reach distances at which multiple
events may be detected per year ~\cite{whitepaper,EURO,LCGT}.

The sensitivity of interferometric gravitational wave observatories is limited by the
fundamental noise sources inherent in the instrument. In advanced LIGO, thermal noise
from the internal degrees of freedom of the interferometer test masses is expected to be
the limiting noise source in the middle frequency range ($\sim 30$ -- 500 Hz). This is
also the interferometer's most sensitive frequency band. Thus, any additional thermal
noise, such as thermal noise associated with optical coatings, will directly reduce the
number of events that advanced LIGO can detect.

The initial LIGO interferometer uses fused silica for the interferometer  test masses,
the beam splitter, and other optics. Fused silica has been shown to have very low
internal friction~\cite{Fraser,Lunin,highq} and will therefore exhibit very low
(off-resonance) thermal noise. This property, coupled with the fact that high quality,
large, fused silica optics are commercially available,  makes fused silica a natural
choice for the initial interferometer. Sapphire, which has even lower internal
friction~\cite{Sheila,Braginskysbook} (although higher thermoelastic loss) is currently
proposed as the material from which to fabricate the optics for use in advanced
LIGO~\cite{whitepaper}. In addition to lower thermal noise, sapphire offers benefits due
to its superior thermal conductivity, which, in transporting heat from the reflective
surface of the test masses, allows a higher power laser to be used. Fused silica is under
continuing study as a fallback material for advanced LIGO should problems arise in the
development of sapphire optics.

In order to use the test masses as mirrors, optical coatings must be applied to the
surface. To obtain high reflectivities, multi-layer, dielectric coatings are used. Such
coatings consist of alternating layers of two dielectric materials with differing
refractive indices. The number of layers deposited determines the reflectivity. It is
possible to use a number of different dielectric material pairs for reflective coatings,
but it has been found that coatings made with alternating layers of Ta$_2$O$_5$ and
SiO$_2$ give the necessary reflectivity while at the same time satisfying the stringent
limits on optical loss and birefringence required for LIGO~\cite{Camp}. The effect of
these coatings on thermal noise is only now being studied.

The simplest way to predict the thermal noise is to use the Fluctuation-Dissipation
Theorem~\cite{callen}. It states that the thermal noise power spectrum is proportional to
the real part of the mechanical admittance of the test mass. Explicitly
\begin{equation}\label{FDT} S_x \left(f\right) = \frac{k_B T}{\pi^2 f^2}
\realpart{{\mathbb Y}(f)},
\end{equation}
where $S_x$ is the spectral density of the thermally induced fluctuations of the test
mass surface read by the interferometer,  $T$ is the temperature, and $f$ is the
frequency of the fluctuations. The quantity ${\mathbb Y}(f)$ is the mechanical admittance
of the test mass to a cyclic pressure distribution having the same form as the
interferometer beam intensity profile~\cite{Levin}. For LIGO, the proposed beam profile
is Gaussian. The real part of the admittance, used in the theorem, can be written in
terms of the mechanical loss angle, $\phi\sub{readout}$, of the test mass response to the
applied cyclic Gaussian pressure distribution.  To calculate the thermal noise we must
therefore obtain $\phi\sub{readout}$.

The loss angle $\phi\sub{readout}$ depends both on the distribution of losses in the test
mass and on the shape of the deformation of the test mass in response to the applied
pressure.  If the distribution of losses in the test mass were homogeneous, the loss
angle $\phi\sub{readout}$ would be independent of the deformation of the test mass.  In
that case, one could obtain $\phi\sub{readout}$ by measuring the loss angle associated
with a resonant mode of the test mass, $\phi=1/Q$, where $Q$ is the quality factor of a
resonant mode. However, when the distribution of mechanical losses in the test mass is
not homogeneous, this approach does not work.

One way of obtaining $\phi\sub{readout}$ would be to measure it directly.  This would
involve applying a cyclic Gaussian pressure distribution to the test mass face and
measuring the phase lag of the response.
But such an experiment presents several insuperable technical difficulties and is useful
mainly as a thought experiment, in which interpretation of the result would be simple.

In this paper, we give the results of another kind of experiment whose results allow us
to calculate $\phi\sub{readout}$ using elasticity theory.  The measurement process is
relatively straightforward: we compare the quality factor, $Q$, of vibrations of an
uncoated sample of fused silica to the quality factor when a coating has been applied. In
order to make the effect easier to measure, and to improve the accuracy of the
measurements, we used thin pieces of fused silica rather than the relatively thick
mirrors used in LIGO. Our measurements show a significant reduction of the $Q$ due to
mechanical loss associated with the coating.

In choosing to make the measurements easy to carry out, we necessarily complicated the
interpretation of the results. Scaling from the results of our measurements to the
prediction of $\phi\sub{readout}$ takes some work. In Section~\ref{Theory} we describe
the relationship between the measured loss angle $\phi_\|$ and the readout loss angle
$\phi\sub{readout}$ . We are then able, in Section~\ref{Method}, to more succinctly
explain the methodology and details of our measurement process.  The results of the
measurements are described in Section~\ref{Results}. The implications for LIGO are
discussed Section~\ref{Implications}, and a program of future work is described in
Section~\ref{FutureWork}.  These results are complemented by similar results on
\chem{Al_2O_3/Ta_2O_5} coatings deposited on thick fused silica substrates.  Those
results are published in a companion paper~\cite{glasgowcoating}.

\section{Theory}\label{Theory}
\renewcommand{\arraystretch}{1.3}

To use the Fluctuation-Dissipation Theorem, \eqn{FDT}, to predict thermal noise, we need
to calculate the real part of the mechanical admittance of the test mass. The mechanical
admittance of the test mass is defined as
\begin{equation}\label{AdmittanceDefinition}
    {\mathbb Y}(f)\equiv i\,2\pi f \, \frac{ x(f)}{F}
\end{equation}
where $F$ is the (real) amplitude of a cyclic pressure distribution applied to the test
mass at frequency $f$, and $x(f)$ is the (complex) amplitude of the steady state
response. Choosing the appropriate pressure distribution with which to excite the test
mass constitutes the first step in the calculation. Levin \cite{Levin} has argued that in
calculating the thermal noise read by an interferometer, the appropriate pressure
distribution has the same profile as the laser beam intensity and should be applied to
the test mass face (in the same position and orientation as the beam). In the case of
initial LIGO the laser beam has a Gaussian intensity distribution, and a Gaussian beam
profile is also proposed for advanced LIGO. The corresponding cyclic pressure
distribution is
\begin{equation}\label{DynamicPressureDistribution}
p(\vec{r},t) = p(r,t) = \frac{2F}{ \pi w^2}\,
\exp\!\left(\frac{-2r^2}{w^2}\right)\,\sin{\!(2\pi f t)},
\end{equation}
where $\vec{r}$ is a point on the test mass surface, $r=\abs{\vec{r}}$, $f$ is the
frequency of interest, and $w$ is the field amplitude radius of the laser beam. (At the
radius $w$, the light intensity is $1/e^2$ of maximum).  To simplify the calculation of
the response $x(f)$, we make use of the fact that the beam radius is considerably smaller
than the test mass radius, and approximate the test mass by an infinite half-space.  This
allows us to ignore boundary conditions everywhere except on the face of the test mass.
For the case of homogeneous loss, Liu and Thorne~\cite{ThorneLiu} have shown that this
approximation leads to an overestimate of the thermal noise, but that for a test mass of
radius 14~cm, the error is about 30\% or less for beam field amplitude radii $w$ up to
6~cm.

To calculate the real part of the admittance we follow Levin and rewrite it in the form
\begin{equation}\label{Yofphi}
\realpart{{\mathbb Y}(f)}=\frac{4\pi f \, U(f)}{F^{\,2}}\,\phi,
\end{equation}
where $U\left(f\right)$ is the maximum elastic energy stored in the test mass as a result
of the excitation, and $\phi$ is the loss angle of the response. \Eqn{Yofphi} holds at
frequencies far below the first resonance of the test mass, provided $\phi\ll 1$, and is
obtained as follows. Under the conditions stated
\begin{equation}\label{InteractionHamiltonian} U(f) = \frac{1}{2} \, \abs{F}\,
\abs{x(f)},
\end{equation}
and the response $x(f)$ to the excitation is
\begin{equation}\label{SmallPhiLowfApprox}
x(f)=\abs{x(f)}\exp(-i\phi) \approx \abs{x(f)}\,(1-i\phi).
\end{equation}
Substituting Eqs.~(\ref{InteractionHamiltonian})~and~(\ref{SmallPhiLowfApprox}) into
\eqn{AdmittanceDefinition} and taking the real part yields \eqn{Yofphi}.

The strategy is then to calculate $U(f)$ and $\phi$ under the pressure distribution in
\eqn{DynamicPressureDistribution}.  Calculation of the loss angle $\phi$ requires some care since
the loss angle is specific to the applied force distribution and to the associated
deformation. If the material properties or intrinsic sources of loss are  not isotropic
and homogeneous throughout the sample, different deformations will exhibit different loss
angles.  Since interferometer test masses do have inhomegeneous loss due to the
dielectric coating on the front surface, the calculation of thermal noise depends on
obtaining the value of the loss angle associated with precisely the response to the
pressure distribution given in \eqn{DynamicPressureDistribution}. Throughout this paper
we will assume that losses in the substrate are always homogeneous and isotropic and that
the source of inhomogeneous and anisotropic loss is the coating.

The loss angle $\phi\equiv\phi\sub{readout}$ associated with the Gaussian pressure
distribution can be written as a weighted sum of coating and substrate losses. If the
loss in the coating is homogeneous and isotropic within the coating (but still different
from that of the substrate) we can write
\begin{equation}\label{PhiAddition}
\phi\sub{readout} =\frac{1}{U}
(U\sub{substrate}\,\phi\sub{substrate} +U\sub{coating}\,\phi\sub{coating})
\end{equation}
where $U$ is the maximum elastic energy stored in the sample as a consequence of the
applied pressure, $U\sub{substrate}$ is the portion of the energy stored in the
substrate, $U\sub{coating}$ is the portion of the energy stored in the coating,
$\phi\sub{substrate}$ is the loss angle of the substrate, and $\phi\sub{coating}$ is the
loss angle of the coating. To simplify the calculation of the energies, we make use of
the fact that the frequencies where thermal noise dominates interferometer noise budgets
are far below the first resonances of the test masses. Thus, the shape of the response of
the test mass to a cyclic Gaussian pressure distribution of frequency $f$ is well
approximated by the response to an identical Gaussian pressure distribution that is
constant in time.  Thus, to a good approximation, $U$, $U\sub{substrate}$, and
$U\sub{coating}$ can be calculated from the deformation associated with the {\em static}
Gaussian pressure distribution
\begin{equation}\label{PressureDistribution}
p(r) = \frac{2F}{ \pi w^2}\,
\exp\!\left(\frac{-2r^2}{w^2}\right).
\end{equation}

Since we are in the limit where the coating is very thin compared to the width of the
pressure distribution
\begin{equation}
U\sub{coating} \approx \delta U\,d,
\end{equation}
where $\delta U$ is the energy density stored at the surface, integrated over the
surface, and $d$ is the thickness of the coating.  Similarly, $U\sub{substrate}\approx
U$, giving
\begin{equation}\label{PhiReadoutIsotropic}
\phi\sub{readout}=\phi\sub{substrate}+ \frac{\delta
U\,d}{U}\,\phi\sub{coating}\mbox{\raisebox{-1pt}{.}}
\end{equation}

If the loss angle of the coating is not isotropic, the second term in
\eqn{PhiReadoutIsotropic} must be expanded. Since the coatings have a layer structure, it
may not be accurate to assume that their structural loss is isotropic.  To address the
possible anisotropy of the structural loss we shall use the following model. The energy
density $\rho_{\scriptscriptstyle U}$ of a material that is cyclically deformed will
generally have a number of terms. We shall associate a different (structural) loss angle
with each of these terms. For example, in cylindrical coordinates
\begin{equation}\label{EnergyDensityDef}
\begin{array}{lcl}
\rho_{\scriptscriptstyle U}&=&\rho_{rr} + \rho_{r\theta} + \ldots
\end{array}
\end{equation}
where \vspace{-1ex}
\begin{equation}\label{EnergyDensitiesDefn}
\begin{array}{lcl}
\rho_{rr}       &\equiv&    \frac{1}{2}\,\sigma_{rr}\epsilon_{rr} \\
\rho_{r\theta}  &\equiv&    \frac{1}{2}\,\sigma_{r\theta}\epsilon_{r\theta}\\
                &\vdots&
\end{array}
\end{equation}
where $\sigma_{ij}$ are stresses and $\epsilon_{ij}$ are strains. The associated loss
angles are $\phi_{rr}$, $\phi_{r\theta}$, etc. In this paper we will assume that the loss
angles associated with energy stored in strains parallel to the plane of the coating are
all equal. This assumption is motivated by the observation that many isotropic, amorphous
materials, like fused silica, do not show significantly different quality factors for
many modes even though the relative magnitude of the  various terms in the elastic energy
varies significantly between the modes~\cite{Numata}.  The measurements made at Glasgow
and Stanford Universities further strengthen this
assumption~\cite{glasgowcoating}\,---\,those measurements show no significant variation
of the coating loss as the relative size of the different coating energy terms changes
from mode to mode. Note that since we will always have traction free boundary conditions
for the problems considered here, we shall always have $\epsilon_{rz}=\epsilon_{zr}=0$.
Thus we will have loss angles associated only with the following coating energy density
components
\begin{equation}\label{EnergyComponents}
\left.\begin{array}{lll}
\rho_{{\scriptscriptstyle U\,}\|}'&=&\frac{1}{2}
(\epsilon_{rr}'\sigma_{rr}' +
\epsilon_{\theta\theta}'\sigma_{\theta\theta}' + \epsilon_{r\theta}'\sigma_{r\theta}')\\
\rho_{{\scriptscriptstyle U}\bot}'&=&\frac{1}{2}\,
\epsilon_{zz}'\sigma_{zz}'\\
\end{array}\right.
\end{equation}
where $\epsilon_{ij}'$ are the strains and $\sigma_{ij}'$ are the stresses in the
coating. We define the loss angle associated with the energy density in parallel coating
strains $\rho_{{\scriptscriptstyle U}\|}$\,, as $\phi_\|$, and the loss angle associated
with the density of energy in perpendicular coating strains, $\rho_{{\scriptscriptstyle
U}\bot}'$, as $\phi_\perp$.  The components of the energy density in
\eqn{EnergyComponents} integrated over the surface of the (half-infinite) test mass are
\begin{equation}\label{dUdefn}
\left.\begin{array}{lll}
\delta U_{\|}   &=& \int_{S} \rho_{{\scriptscriptstyle U\,}\|}'\,d^{\,2}r\\
\delta U_{\bot} &=& \int_{S} \rho_{{\scriptscriptstyle U}\bot}'\,d^{\,2}r.
\end{array}\right.
\vspace{0.2ex}
\end{equation}
So that finally, to account for the anisotropic layer structure of the coating,
\eqn{PhiReadoutIsotropic} is replaced by
\begin{equation}\label{PhiReadout}
\phi\sub{readout}=\phi\sub{substrate}+ \frac{\delta
U_{\|}\,d}{U}\,\phi\sub{\|} + \frac{\delta U_{\bot}\,d}{U}\,\phi\sub{\bot}.
\end{equation}

To obtain an expression for  $\phi\sub{readout}$ we need to calculate $\delta U_{\|}$,
$\delta U_{\bot}$,  and $U$ for a coated half-infinite test mass subject to the Gaussian
pressure distribution $p(r)$ of Eq.~(\ref{PressureDistribution}). The quantities, $\delta
U_{\|}$ and $\delta U_{\bot}$ involve only the stress and strain in the coating. The
total energy involves the stress and strain throughout the substrate
\begin{equation}\label{TotalEnergyDefn}
U=\pi \int_0^\infty dz \int_0^\infty r\,dr\,
(\epsilon_{rr}\sigma_{rr}+\epsilon_{\theta\theta}\sigma_{\theta\theta}+
\epsilon_{zz}\sigma_{zz} + 2\epsilon_{rz}\sigma_{rz}),
\end{equation}
where $\epsilon_{ij}$ are the strains and $\sigma_{ij}$ the stresses in the substrate. To
obtain the stresses and strains in the coating and in the substrate we must solve the
axially symmetric equations of elasticity for the coated half-infinite test mass subject
to the pressure distribution $p(r)$. The general solution to these equations for an
uncoated half-infinite test mass is given by Bondu {\it et al.}~\cite{Bondu} (with
corrections by Liu and Thorne~\cite{ThorneLiu}).

Because the coating is thin, we can, to a good approximation, ignore its presence in the
solution of the elastic equations for the substrate. The strains in the coating should
also not vary greatly as a function of depth within the coating, and we shall approximate
them as being constant.  Due to axial symmetry, $\epsilon_{r\theta}=\epsilon_{\theta
r}=\epsilon_{r\theta}'=\epsilon_{\theta r}'=0$. Due to traction-free boundary conditions,
$\epsilon_{rz}'=\epsilon_{zr}'=0$ at the coating surface, and the same must therefore
hold (to leading order) for the entire coating. This approximation is valid, provided the
Poisson's ratio of the coating is not very different from that of the substrate. To
obtain the non-zero stresses and strains in the coating ($\epsilon_{rr}'$,
$\epsilon_{\theta\theta}'$, and $\epsilon_{zz}'$) we note that since the coating is
constrained tangentially by the surface of the substrate, the coating must have the same
tangential strains ($\epsilon_{rr}'$ and $\epsilon_{\theta\theta}'$) as the surface of
the substrate. Also, the coating sees the same perpendicular pressure distribution
($\sigma_{zz}'$) as the surface of the substrate.  These conditions, which represent
reasonably good approximations for the case of a thin coating, allow us to calculate all
the coating stresses and strains in terms of the stresses and strains in the surface of
the substrate. See Appendix~\ref{CoatingStressesAndStrainsAppendix} for the details of
this calculation.

Using the solutions for $\epsilon_{ij}'$, $\sigma_{ij}'$, $\epsilon_{ij}$, and
$\sigma_{ij}$ derived in Appendix~\ref{CoatingStressesAndStrainsAppendix}, and
substituting into Eqs.~(\ref{EnergyComponents}-\ref{TotalEnergyDefn}), we obtain the
required quantities
\begin{eqnarray}
U&=&\frac{F^2(1-\sigma^2)}{2\sqrt{\pi}\,wY}.\label{TotalEnergy}\\
\delta U_{\|}/U  &=& \frac{1}{\sqrt{\pi}\,w}\,
\frac{Y'(1+\sigma)(1-2\sigma)^2+Y\sigma'(1+\sigma')(1-2\sigma)}
{Y(1+\sigma')(1-\sigma')(1-\sigma)}\label{dUpara}\\
\delta U_{\bot}/U &=&\frac{1}{\sqrt{\pi}\,w}\,
\frac{Y(1+\sigma')(1-2\sigma')-Y'\sigma'(1+\sigma)(1-2\sigma)}
{Y'(1-\sigma')(1+\sigma)(1-\sigma)},\label{dUperp}\\
\end{eqnarray}
where $Y$ and $\sigma$ are the Young's modulus and Poisson's ratio of the substrate, and
$Y'$ and $\sigma'$ are the Young's modulus and Poisson's ratio of the coating. Thus, from
\eqn{PhiReadout}
\begin{multline}\label{PhiReadoutResult}
\phi\sub{readout}=\phi\sub{substrate}+\frac{1}{\sqrt{\pi}}\,\frac{d}{w}\,
\left(
\frac{Y'(1+\sigma)(1-2\sigma)^2+Y\sigma'(1+\sigma')(1-2\sigma)}
{Y(1+\sigma')(1-\sigma')(1-\sigma)}\:\phi_{\|} \; +
\right.\\
\left.
\frac{Y(1+\sigma')(1-2\sigma')-Y'\sigma'(1+\sigma)(1-2\sigma)}
{Y'(1-\sigma')(1+\sigma)(1-\sigma)}\:\phi_{\bot}\right).
\end{multline}

Substituting Eqs.~(\ref{TotalEnergy})~and~(\ref{PhiReadoutResult}) into \eqn{Yofphi} and
substituting the result into the Fluctuation-Dissipation Theorem, \eqn{FDT}, gives the
power spectral density of interferometer test mass displacement thermal noise as
\begin{multline}\label{ThermalNoiseResult}
S_x(f)=\frac{2k_BT}{\pi^{3/2}f}\,\frac{1-\sigma^2}{wY}
\bigg\{\phi\sub{substrate} +
\frac{1}{\sqrt{\pi}}\,\frac{d}{w}\,\frac{1}{Y Y'(1-{\sigma'}^2)(1-\sigma^2)}
\\
\Big[ {Y'}^2(1+\sigma)^2(1-2\sigma)^2\,\phi_\| +
YY'\sigma'(1+\sigma)(1+\sigma')(1-2\sigma)\,(\phi_\|-\phi_\bot)
+Y^2(1+\sigma')^2(1-2\sigma')\,\phi_\bot
\Big] \bigg\} .
\end{multline}
Equation~(\ref{ThermalNoiseResult}) is valid provided that most of the loss at the coated
surface occurs in the coating materials themselves and is not due to interfacial rubbing
between the coating and the substrate, or to rubbing between the coating layers. If a
large proportion of the loss is due to rubbing, the coating-induced thermal noise will
not be proportional to the coating thickness as indicated in
Eq.~(\ref{ThermalNoiseResult}). Rather, it may be proportional to the number of layers
and may be very dependent on the substrate preparation.

The limits of \eqn{ThermalNoiseResult} agree with previous results. In the limit that
$\phi_\|=\phi_\bot$, the $YY'$ term disappears and the result agrees with the result of
Nakagawa who has solved the problem for that case by a different method \cite{NorioNew}.
The limit of \eqn{ThermalNoiseResult} in the case $Y'=Y$, $\sigma'=\sigma$, and
$\phi_\bot=\phi_\|$ agrees with the result obtained previously
\cite{NakagawaAndGretarsson} \begin{equation}\label{ThNoiseOldResult}
S_x(f)=\frac{2k_BT}{\pi^{3/2}f}\,\frac{1-\sigma^2}{wY}
\left\{ \phi\sub{substrate}+\frac{2}{\sqrt{\pi}}\,\frac{(1-2\sigma)}{(1-\sigma)}\,
\frac{d}{w}\,\phi_\|\right\}.
\end{equation}

For the case of fused silica or sapphire substrates coated with alternating layers of
\chem{Ta_2O_5} and \chem{SiO_2}, the Poisson's ratio of the coating may be small enough
($\lsim 0.25$) that for likely values of the other parameters, \eqn{ThermalNoiseResult}
is reasonably approximated (within about 30\%) by the result obtained by setting
$\sigma=\sigma'=0$ \begin{equation}\label{ZeroPoissonsRatio}
S_x(f)=\frac{2k_BT}{\pi^{3/2}f}\,\frac{1}{wY}
\left\{ \phi\sub{substrate}+\frac{1}{\sqrt{\pi}}\,\frac{d}{w}\,
\left(\frac{Y'}{Y}\,\phi_{\|} +
\frac{Y}{Y'}\,\phi_\bot\right)\right\}.
\end{equation}
\Eqn{ZeroPoissonsRatio} highlights the significant elements of \eqn{ThermalNoiseResult}.
It shows that in order to estimate the thermal noise performance of a particular coating,
we must know all of $Y$, $Y'$, $\phi_\|$, and  $\phi_\perp$.  It also shows that if
$\phi_\|\approx\phi_\bot$, then the lowest coating-induced thermal noise occurs when the
Young's modulus of the coating is matched to that of the substrate. If $Y'
\not\approx Y$, one of $\phi_\|$ or $\phi_\bot$ will be emphasized and the other
de-emphasized.  This is particularly worrisome for coatings on sapphire substrates, whose
high Young's modulus means that for most coatings $\phi_\bot$ is likely to be the main
contributor to the coating thermal noise.  \Secn{Method} describes ringdown experiments
on coated samples in order to determine~$\phi_\|$.   Unfortunately we don't obtain
$\phi_\bot$ from ringdown experiments of samples with coatings on the surface. Since the
coatings experience free boundary conditions they are not greatly compressed
perpendicular to the surface (there will be some small amount of compression due to
Poisson ratio effects). Therefore $\phi_\bot$ cannot be easily measured in such
experiments, and no measurement of $\phi_\bot$ exists at the present time. Because of
this, we can only obtain very rough estimates of the coating-induced thermal noise.  We
will set $\phi_\bot=\phi_\|$, but the accuracy of our thermal noise estimates will remain
unknown until $\phi_\bot$ is measured.

\renewcommand{\arraystretch}{1}

\section{Method}\label{Method}
\renewcommand{\arraystretch}{1.3}
In order to estimate the coating loss component $\phi_\|$, we made measurements of the
loss angles of fused silica samples with and without the
\chem{Ta_2O_5/SiO_2} high reflective coating used in LIGO. A standard way of determining
the loss angle at the frequency of a particular resonant mode is to measure its ringdown
time, $\tau_n$. This allows the calculation of the mode's quality factor $Q$, through
\begin{equation}
Q \equiv \pi f_n \tau_n,
\end{equation}
where $f_n$ is the frequency of the resonant mode. The loss angle at the resonance
frequency is the inverse of the mode's quality factor \begin{equation}\label{Qdef}
\phi\left(f_n\right) = 1/Q.
\end{equation}

Because of the free boundary conditions no energy is stored in strains having
perpendicular components.  The loss angle $\phi\sub{coated}$ of a resonating sample after
coating is therefore related to the loss angle $\phi\sub{uncoated}$ of the same sample
before coating by
\begin{equation}\label{RingdownPhi}
\phi\sub{coated}=\phi\sub{uncoated}
+ \frac{\delta \tilde{U}_\| \,d}{\tilde{U}}\, \phi_\|
\end{equation}
where $\tilde{U}$ is the energy stored in the resonance. Similarly, as in \secn{Theory},
the quantity $\delta\tilde{U}_\|$ is the resonance energy stored in strains having no
component perpendicular to the surface
\begin{equation}\label{RingdowndU}
\delta\tilde{U}_\| =  \int_{S}d^{\,2}r\,
\sum_{i,j\,\ne z} \rho'_{ij},
\end{equation}
where ${S}$ is the coated surface of the sample, $z$ is the direction perpendicular to
the surface, and
\begin{equation}
\rho'_{ij} = \frac{1}{2}\,\epsilon_{ij}' \,\sigma_{ij}'.
\end{equation}
Just as in \secn{Theory}, $\phi_\|$ in \eqn{RingdownPhi} is the loss angle associated
with energy stored in strains in the plane of the coating. Because we assume that all
in-plane loss angles are identical, the loss angle $\phi_\|$ is the same $\phi_\|$ as in
\secn{Theory}, and once measured, can be substituted directly into \eqn{ThermalNoiseResult}.

For each sample resonance that was found, $\phi\sub{coated}$ and $\phi\sub{uncoated}$
were measured by recording the $Q$ with and without an optical coating, respectively. The
quantity $(\delta\tilde{U}_\|\,d/ \tilde{U})$ was then calculated either numerically or
analytically, allowing \eqn{RingdownPhi} to be solved for $\phi_\|$ based on the measured
values of $\phi\sub{coated}$ and $\phi\sub{uncoated}$.  The resulting value for $\phi_\|$
was then substituted into \eqn{ThermalNoiseResult} to obtain an interferometer thermal
noise estimate.

\begin{figure}
\includegraphics[width=12cm]{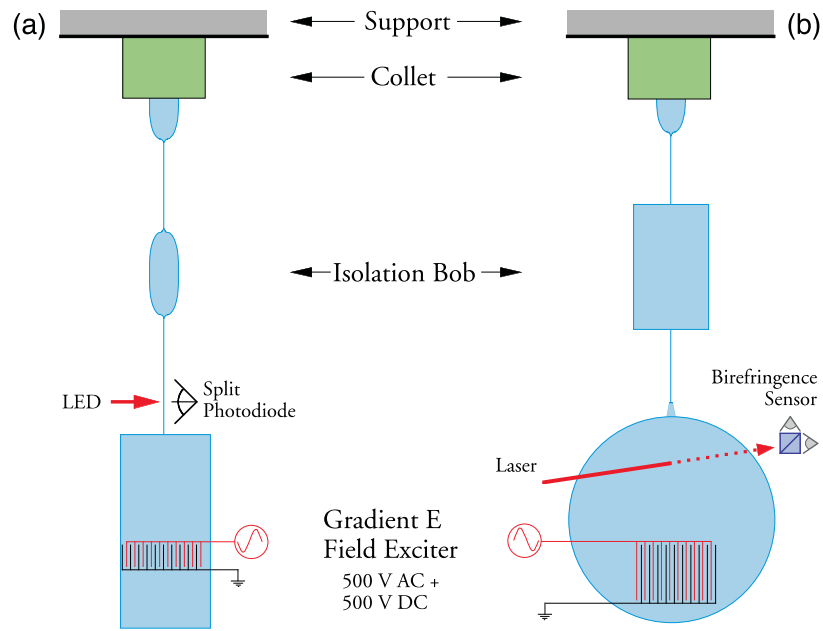}\vspace{2ex}
\caption{
{\bf (a)}~The suspended microscope slide and exciter.~~{\bf (b)}~The suspended disk and
exciter.  In both {\bf (a)} and {\bf (b)}, the entire structure below the steel collet is
fused silica.}\label{fig:sample}
\end{figure}

In order to reduce systematic errors in the $Q$ measurements, we took a number of steps
to reduce excess loss (technical sources of loss, extrinsic to the
sample)~\cite{YingleiPaper,Gretarsson}.  All the $Q$'s were measured in a vacuum space
pumped down to at least $1 \times 10^{-5}$~torr, and more typically $2 \times
10^{-6}$~torr. This reduced mechanical loss from gas damping. During the $Q$
measurements, the samples were hung below a monolithic silica suspension made by
alternating a massive bob of silica with thin, compliant, silica fibers.  The suspensions
and samples are shown in Fig.~\ref{fig:sample}.  (The suspension is of the same style
used previously in~\cite{highq},~\cite{Gretarsson}, and~\cite{amaldi}.) The piece of
fused silica rod at the top of the suspension is held in a collet which is rigidly
connected to the underside of a thick aluminum plate supported by three aluminum columns.
Between the piece of rod held in the collet and the sample was a single fused silica
isolation bob. Its function was to stop vibrations from traveling between the sample and
the aluminum optical table from which it was suspended. The size chosen for the isolation
bob depended on the sample, with the heavier sample requiring a larger bob. The two
fibers in the suspension were monolithically pulled out of the neighboring parts using a
H$_2$-O$_2$ torch. These fibers had a typical diameter of roughly 100-200~$\mu$m. The
normal modes of the sample were excited using a comb capacitor~\cite{Cadez}. This exciter
was made from two copper wires sheathed with Teflon, each having a total diameter of
about 1/2~mm.  The two wires were then wrapped around a ground plane and placed about
1~mm from the face of the sample. Special care was taken to ensure that the exciter and
the sample did not touch at any point. The position of the exciter is shown in
Fig~\ref{fig:sample}. Alternating wires of the comb capacitor were given a 500~V DC
voltage while the other wires were held at ground to induce a polarization in the glass
sample. To reduce any eddy current damping~\cite{YingleiPaper} and to reduce the
probability that polarized dust could span the gap between the sample and the exciter,
the exciter was always kept more than 1/2~mm away from the sample.  An AC voltage at a
resonance frequency of the sample was then added to the DC voltage to excite the
corresponding mode. Once the mode had been excited (``rung up'') to an amplitude where it
could be seen clearly above the noise, both the AC and DC voltages were removed and both
exciter wires were held at ground.  The sample was then allowed to ring down freely.

The amplitude of excitation in the sample was read out using a birefringence
sensor~\cite{Startin,AA} or (in the earliest measurements) by a shadow sensor. For the
birefringence sensor, a linearly polarized beam is passed through the sample at or near a
node of the resonant mode under study. Modally generated stress at the node induces
birefringence in the glass, which couples a small amount of the light into the orthogonal
polarization, phase shifted by $\pi/2$. Thus, the light exiting the sample is slightly
elliptically polarized. The beam is then passed through a $\lambda/4$ wave-plate aligned
with the initial polarization.  This brings the phases of the two orthogonal polarization
components together, converting the elliptically polarized light to a linear polarization
that is rotated slightly compared with the initial polarization. The rotation angle is
(to first order) proportional to the modal strain, and is measured by splitting the beam
with a polarizing beamsplitter and monitoring the relative intensity of light in the two
channels. This was done with two identical photodiodes and a differencing
current-to-voltage amplifier. The output voltage oscillates sinusoidally at the resonant
frequency in proportion to the modally induced strain. This signal is sent to a lock-in
amplifier to demodulate it to a lower frequency, and the data is collected on a PC. The
ringdown time $\tau_n$ was obtained by fitting the acquired signal to a damped sinusoid,
or by extracting the envelope of the decay.  Both approaches yielded the same results,
although the accuracy of the former was less sensitive to corruption from noise. A
schematic drawing of the optical readout system is shown in Fig.~\ref{fig:readout}.

For the shadow sensor, an LED is used to cast the shadow of the fused silica suspension
fiber onto a split photodiode.  The LED/diode pair is positioned close to where the
suspension fiber is welded to the edge of the sample.  The fiber near the weld point will
faithfully follow the motion of the edge of the sample. As the sample resonates, the
fiber's shadow moves back and forth on the photodiode at the same frequency. The amount
of light falling on each half of the split photodiode changes proportionally. The
currents from each half of the photodiode are then compared with a differential
current-to-voltage amplifier as in the case of the birefringence sensor. The data
acquisition and analysis were identical for both sensors.

For relatively rigid but transparent samples like the ones used here, the birefringence
sensor is significantly more sensitive and much easier to use than the shadow sensor. The
shadow sensor is better suited to more compliant samples.  In both cases however, the
dominant sources of broadband noise were laser noise and noise from the differential
amplifier.

The samples were coated by Research Electro-Optics Corporation (REO) of Boulder,
Colorado, USA. The coating was a dielectric optical coating consisting of alternating
layers of SiO$_2$ and Ta$_2$O$_5$.  The coating was laid down using Argon ion beam
sputtering, followed by annealing at 450$^\circ$C. We chose to examine this particular
type of coating because it is the one used on the initial LIGO mirrors that are currently
installed at the LIGO sites.  This coating is also the type currently proposed for
advanced LIGO optics.

The first samples we studied were three rectangular prisms in the shape of microscope
slides (7.6~cm $\times$ 2.5~cm $\times$ 0.1~cm) made of Suprasil 2 brand fused silica
from Heraeus Quartzglas GmbH of Hannau, Germany. The surface of these samples was treated
with a commercial polish to a scratch/dig specification of 80/50.  There was no
specification on the overall flatness or the surface figure.  Two of the three slides
(Slide A and Slide B) were coated on both sides with a reflective
\chem{Ta_2O_5}/\chem{SiO_2} coating of 3\% transmittance for normally incident,
$1~\micron$ wavelength light. The third slide (Slide C) was left uncoated as a control.
Slide~A was suspended from a corner, which had remained uncoated due to being supported
at those points during coating. Therefore, welding the suspension fiber to the corner did
not induce visible damage to the coating. Slide~B on the other hand was suspended from
the center of one of its short edges. During the weld, the coating near the suspension
point was visibly damaged in a small crescent shaped of radius ~2~mm surrounding the
suspension point. This damaged region was etched off using hydrofluoric (HF) acid.
Table~\ref{table:slide} shows the modes and quality factors for which $Q$'s were
repeatably measured.  A preliminary version of these results was reported at the Third
Edoardo Amaldi Conference On Gravitational Waves~\cite{amaldi}.

\begin{figure}
\includegraphics[width=8cm]{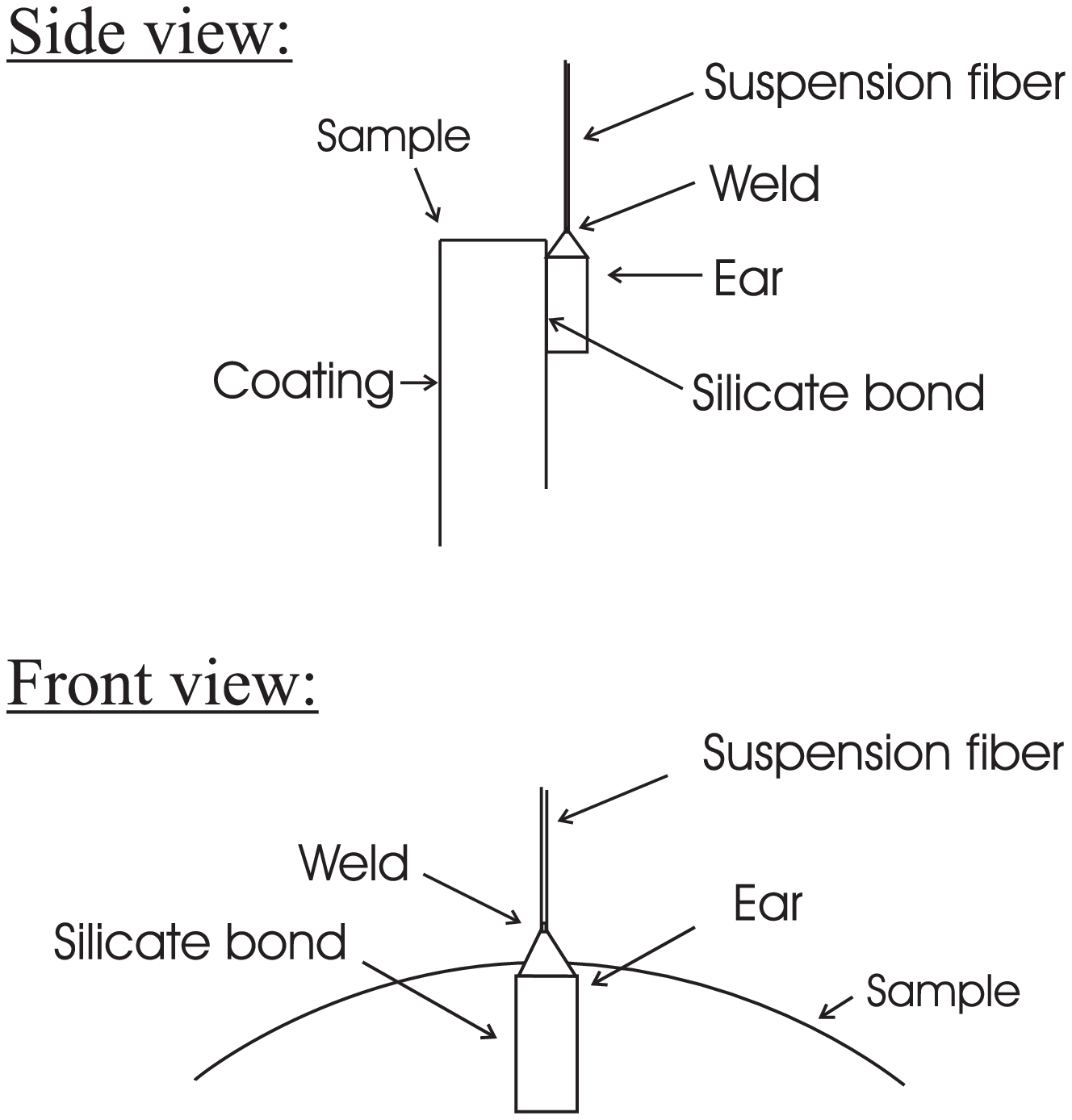}\vspace{2ex}
\caption{
Details of the attachment point. The suspension fiber is welded to the top of the ear.
The ear is in turn silicate bonded along one of its flat faces to the uncoated side of
the sample.}\label{fig:ear}
\end{figure}

\begin{figure}
\includegraphics[width=12cm]{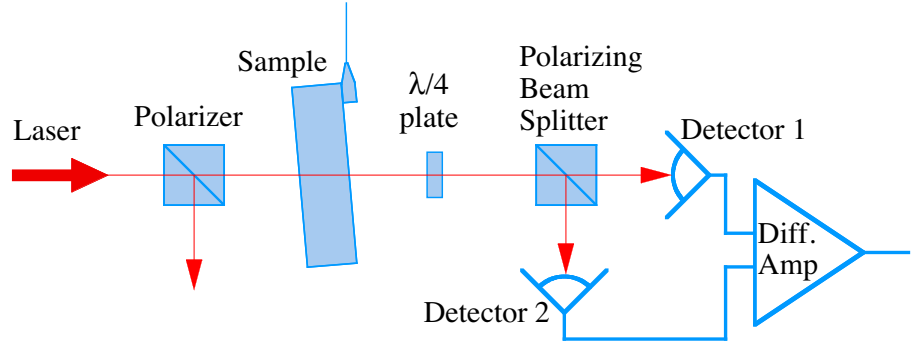}\vspace{2ex}
\caption{Layout of the birefringence sensor. }\label{fig:readout}
\end{figure}

After measuring the $Q$'s of the slides, we obtained from Zygo Corporation of Middlefield
Connecticut a disk of Dynasil brand fused silica, 164.85~mm in diameter and 19.00~mm
thick.  In an effort to determine the effect of surface preparation on the loss due to
optical coatings, this sample was made with strict specification on surface flatness,
scratch/dig, and surface roughness. The coated surface had a surface flatness less than
$\lambda/20$ ($\lambda=633~\mathrm{nm}$), a scratch/dig of 60/40, and a surface roughness
less than $4$~\AA~rms.  The back surface had a surface flatness less than $\lambda/6$, a
scratch/dig of 60/40, and a surface roughness less than $4$~\AA~rms. These specifications
are nearly as stringent as the actual requirements for LIGO mirrors.  To avoid destroying
the surface with welding, an ``ear'' of fused silica was bonded onto the back surface
using hydroxy catalysis bonding (silicate bonding) \cite{silicatebond}. This ear is
shaped like a rectangular block with a pyramid on one face.  One face of the block is
bonded to the sample, so that the tip of the pyramid faces radially. This allows the
monolithic suspension to be welded with a torch to the tip of the pyramid without heating
the sample very much. See Fig.~\ref{fig:ear}. Once hung, the $Q$ of the sample was
measured using the birefringence readout.

Due to the thickness of this sample (required to meet the flatness specification), only
one normal mode had a frequency below 5~kHz.  The useful bandwidth of the high voltage
amplifier that was used to drive the exciter is about 5~kHz, so measurements were
possible only on this mode.  This was the ``butterfly'' mode, with two radial nodal lines
($\ell = 2$) and  no circumferential nodal lines (n=0)~\cite{Blevins}.


After measuring the $Q$ of this uncoated sample, it was sent to REO to be coated.  It
received a high reflective (HR) coating on one side having 1~ppm transmittance and
optimized for a $45^\circ$ angle of incidence. The sample was then rehung and the $Q$
remeasured.  As can be seen from Table~\ref{table:thickdisk}, the coating caused a
significant reduction in the quality factor.  To rule out possible excess loss due to the
suspension, the sample was then removed and again rehung. During this hanging attempt
(between successful hangings numbers 3 and 4 in Table~\ref{table:thickdisk}), the
isolation bob fell and sheared off the bonded ear. The bond did not give; rather,
material from the sample pulled out along with the ear. A second ear was re-bonded at
180$^\circ$ to the original ear. Unfortunately, this ear was also sheared off in the same
way during the attempt to suspend the sample. This time the source of the break occurred
along the bonded surface, although some of the substrate pulled away as well. Finally, a
third attempt succeeded with an ear bonded at $90^\circ$ to the original ear (hanging
number 4 in Table~\ref{table:thickdisk}). Despite the broken ears, the quality factor of
the coated disk did not change significantly.  The results of all $Q$ measurements on the
disk are shown in Table~\ref{table:thickdisk}.

Since it is difficult in any measurement of high $Q$'s to completely eliminate the
extrinsic, technical sources of loss (excess loss), the quality factors measured for a
given sample varied slightly from mode to mode or within a single mode between different
hangings. Since excess loss always acts to reduce the measured $Q$, the best indicator of
the true internal friction of a sample is the quality factor of the highest $Q$ mode over
all modes and hangings. The spread of measured $Q$'s within single hangings was
relatively small. For example, the three $Q$'s measured in hanging number two (sample
uncoated) were all between $3.1 \times 10^6$ and $2.8 \times 10^6$. The twelve $Q$'s
measured in hanging three (sample coated) were all within $1.28\times 10^6$ to $1.09
\times 10^6$. As can be seen from Tables~\ref{table:slide}~and~\ref{table:thickdisk}, the
measured $Q$'s also did not vary much between modes or hangings, nor between samples in
the case of the two coated slides. The reproducibility of the $Q$'s of the disk argues
strongly that neither the silicate-bonded ear nor the broken ears affected the loss of
the sample.  The range of measured $Q$'s for nominally similar situations is indicative
of the level of the variable excess loss.  Thus, for all our samples, the large
difference in $Q$ between the coated and the uncoated measurements must be due to the
coating, and not to statistical variation, excess loss, nor, in the case of the disk, to
the broken ears.

\section{Results}\label{Results}
\renewcommand{\arraystretch}{1.3}
Using the procedure described in Sec.~\ref{Method}, we obtained $Q$ values from both the
slides and the thick disk.  To calculate $\phi_\|$ from the measured $Q$'s we need to
know the value of $\delta \tilde{U}_\|\,d/U$ for each measured mode of the samples. For
transverse bending of the slides, the strain is approximately
\begin{equation}\label{SlideStrain}
\epsilon_{ij}(\vec{r}) = \left\{
\begin{array}{cl}
\ddd{u_y(z)}{z}\,y & \quad i=j=z\\
0                  & \quad \mbox{otherwise}\\
\end{array}\right.
\end{equation}
where $z$ is the coordinate in the slides' longest dimension, $y$ is the coordinate in
the slides' shortest dimension with the origin in the center plane of the slide, and
$u_y(z)$ is the transverse displacement of the center plane of the slides due to the
bending. Displacements in directions other than $y$ are zero for the transverse bending
modes. This gives
\begin{equation}
\left[\frac{\delta \tilde{U}_\|\,d}{\tilde{U}}\right]\sub{slide} = 7.2 \times 10^{-3}\label{eq:muslide}
\end{equation}
for all transverse bending modes of the slides. The butterfly mode of the disk is more
complex, and an analytical expression for strain amplitude $\epsilon(\vec{r})$ was not
found. We made an FEA model of this sample and calculated $U\sub{coating}/U$ numerically.
This resulted in a value of
\begin{equation}
\left[\frac{\delta \tilde{U}_\| \,d}{\tilde{U}}\right]\sub{disk}
\approx\;\left[\frac{U\sub{coating}}{U}
\right]\sub{disk} = 5.3 \times 10^{-3}\label{eq:mudisk}
\end{equation}
for the butterfly mode of the disk.

The quantities needed to calculate $\phi_{\mathrm{coated}}$ from Eq.~(\ref{RingdownPhi})
are shown in Table~\ref{table:params}. Substituting the $Q$ measurements from
Table~\ref{table:slide} into \eqn{Qdef} to get the loss angles, then using
\eqn{eq:muslide} and the values in Table~\ref{table:params} in
\eqn{RingdownPhi} and solving for $\phi_\|$, we get
\begin{equation}
\phi_{\|,\mathrm{slide}} = 4.2 \pm 0.3 \times 10^{-4}.
\end{equation}
Similarly, from \eqn{eq:mudisk} and the disk $Q$'s in Table~\ref{table:thickdisk} we get
\begin{equation}
\phi_{\|,\mathrm{disk}} = 1.0 \pm 0.3 \times 10^{-4}.
\end{equation}

The agreement in order of magnitude between these two measured values for $\phi_{\|}$
sets a scale for coating thermal noise.  This allows us to make rough estimates of the
effect of coating thermal noise on advanced LIGO.  The value of $\phi_{\|}$ for the
polished disk agrees within its uncertainty with the value measured for coating loss by
the Glasgow/Stanford experiment~\cite{glasgowcoating}, despite the use of a different
coating material in that experiment (\chem{Ta_2O_5}/Al$_2$O$_3$ as opposed to
Ta$_2$O$_5$/\chem{SiO_2}). This suggests that the substrate surface-polish, which is of
similar quality on the disk and on the Glasgow/Stanford samples but less good on the
slides, may be an important factor contributing to the loss. Further tests and
comparisons are necessary before definitive conclusions can be drawn on this issue.

\renewcommand{\arraystretch}{1}

\section{Implications}\label{Implications}

Using Eq.~(\ref{ZeroPoissonsRatio}) for the thermal noise due to the coated mirrors, we
can now estimate the thermal noise spectrum of the advanced LIGO interferometer. We
calculated the range of coating thermal noise in the pessimistic case using the
$\phi_\|=4\times 10^{-4}$ (from the slide results) and in the more optimistic case using
$\phi_\|=1 \times 10^{-4}$ (from the disk result).  In both cases, we assumed a beam spot
size of 5.5~cm, which is the maximum obtainable on fused silica when limited by thermal
lensing effects~\cite{SystemsDesignDocument}.  We have extrapolated our results to
sapphire substrates using the known material properties of sapphire, even though we
didn't measure coating loss directly on sapphire. (There have been recent measurements of
$\phi_\|$ for REO coatings deposited on sapphire~\cite{sheilatalk}. Those results are in
rough agreement with the measurements described here.)  As mentioned before, the thermal
noise estimates will be least accurate for sapphire substrates because sapphire coating
thermal noise is likely to be dominated by $\phi_\perp$ which has not been measured. The
Young's modulus of sapphire is considerably higher than both Ta$_2$O$_5$ and SiO$_2$ in
bulk, so it seems likely that the coating Young's modulus is considerably less than
sapphire's. We are aware of a single, preliminary, measurement of $Y'$ for a
Ta$_2$O$_5$/SiO$_2$ coating~\cite{Nazario} which suggests that the Young's modulus of the
coating is roughly equal to that of fused silica.  We know of no measurements of the
coating's Poisson ratio.  For the purposes of estimating coating thermal noise, we will
set the Young's modulus and Poisson ratio of the coating equal to that of fused silica.

\begin{figure}
\includegraphics[width=15cm]{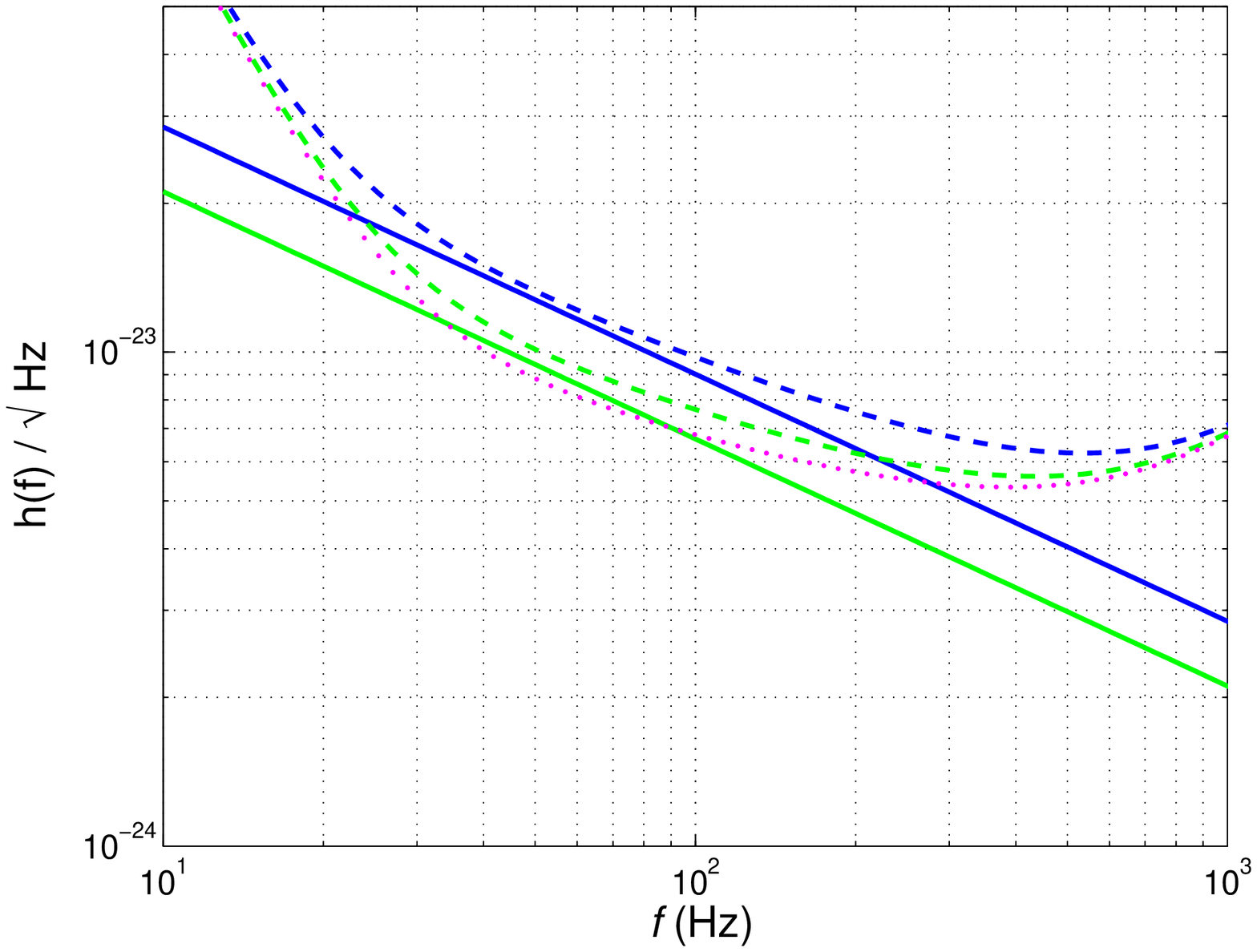}
\caption{Strain spectrum for advanced LIGO with fused silica mirrors.  The solid,
straight lines represent the test mass thermal noise;  the dashed curves show the total
interferometer noise.  The lighter curves were generated using optimistic assumptions
including $\phi_\| = 1 \times 10^{-4}$. The darker curves were generated using
pessimistic assumptions including $\phi_\| = 4 \times 10^{-4}$. The curve shown with
dotted lines is the advanced LIGO noise curve without coating noise as modeled in the
Advanced LIGO System Design document.  In each case, the parameters have been optimized
for binary neutron star inspiral.}\label{fig:fsbench}
\end{figure}

\begin{figure}
\includegraphics[width=15cm]{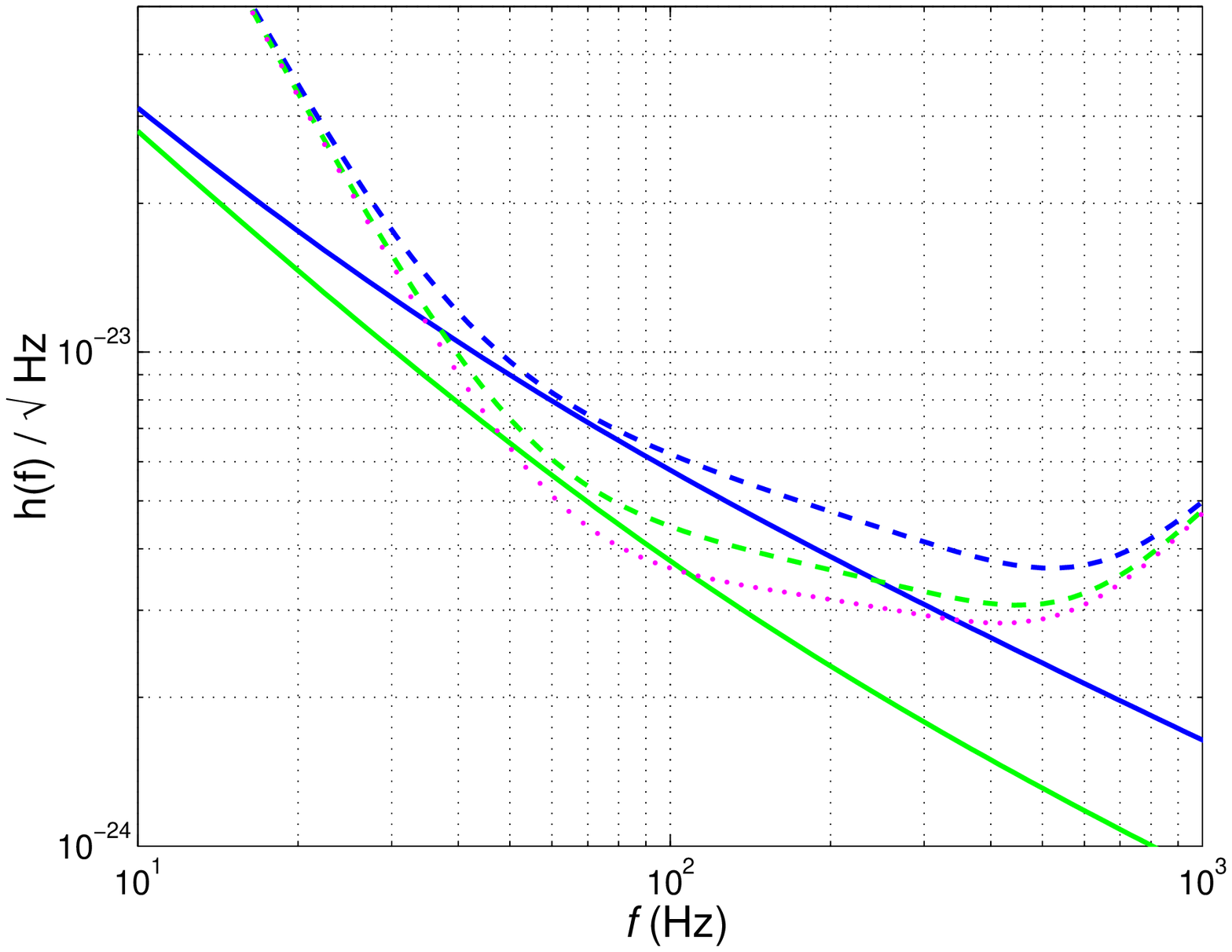}
\caption{Strain spectrum for advanced LIGO with sapphire mirrors.  The
solid, straight lines represent the test mass thermal noise;  the dashed curves show the
total interferometer noise.  The lighter curves were generated using optimistic
assumptions including $\phi_\| = 1 \times 10^{-4}$. The darker curves were generated
using pessimistic assumptions including $\phi_\| = 4 \times 10^{-4}$. The curve shown
with dotted lines is the advanced LIGO noise curve without coating noise as modeled in
the Advanced LIGO System Design document.  In each case, the parameters have been
optimized for binary neutron star inspiral.}\label{fig:sapbench}
\end{figure}

Table~\ref{table:bench} compares the thermal noise estimates for the four cases
considered (optimistic estimates and pessimistic estimates on both fused silica and
sapphire substrates) to the thermal noise estimates when coatings are not taken into
account. The corresponding noise spectra for advanced LIGO are shown in
Figs.~\ref{fig:fsbench}~and~\ref{fig:sapbench}. These were generated using the program
{\tt BENCH\hspace{0.7ex}1.13}~\cite{bench} and show both the total noise and the
contribution from the test mass thermal noise. The curves for the total noise were
generated using the noise models and parameters from the advanced LIGO systems design
document~\cite{SystemsDesignDocument}. The figures show that coating thermal noise is a
significant source of noise in the frequency band $\sim$~30--400~Hz for fused silica test
masses and $\sim$~40--500~Hz for sapphire test masses.

These estimates are only preliminary indications of the level of coating induced thermal
noise.  The largest source of uncertainty in these thermal noise estimates is that no
measurement has been made of $\phi_\perp$. Also, the Young's modulus of the coating
material has not been definitively measured.  The half-infinite test mass approximation
adds extra uncertainty and this estimate needs to be refined by taking the finite size of
the coated test mass into account.  In addition, there remains the possibility that the
loss associated with the different terms in the energy density $\rho_\|$ are not equal as
supposed here.  However, if this were the case, the apparent consistency of the loss
between different modes of the samples measured at Glasgow and
Stanford~\cite{glasgowcoating} would be spurious.

We have also examined the effect of coating thermal noise on the expected sensitivity of
the initial LIGO interferometers that are currently being commissioned.  In initial LIGO,
shot noise will be greater than in advanced LIGO and seismic noise will be significant up
to about 40~Hz.  Due to the higher level of these other noise sources, test mass thermal
noise was not expected to be a large contributor to the total noise~\cite{Abramovici}.
The addition of coating thermal noise raises the overall noise in the most sensitive
frequency band, around 200~Hz, by only 4\% . Thus, coating thermal noise should not
significantly impact the sensitivity of initial LIGO.

In addition to the interferometers used for gravitational wave detection, there are a
number of prototype interferometer within the gravitational wave community.  We have
examined data from one of these\,---\,the 40~m prototype located at Caltech~\cite{40m}.
In this interferometer, the beam spot size was 0.22~cm and the highest $Q$ seen for a
mirror mode was $Q_{\mathrm{max}} = 8.1 \times 10^{6}$~\cite{GillespieRaab}. Using
Eq.~(\ref{ZeroPoissonsRatio}) with $\phi_{\|} = \phi_{\perp} = 1 \times 10^{-4}$ and
$\phi_{\mathrm{substrate}} = 1/Q_{\mathrm{max}}$ in Eq.~(\ref{ZeroPoissonsRatio}) yields
a predicted thermal noise of $\sim 2 \times 10^{-19}$~m/$\sqrt{\mathrm{Hz}}$ at 300~Hz.
This is consistent with Fig.~3 of \cite{40m}. Coating thermal noise is therefore a
possible explanation for the broadband excess noise seen between 300~Hz and 700~Hz. The
effect of coating thermal noise is also being explored in the Glasgow~10~meter prototype,
the Thermal Noise Interferometer (TNI) at Caltech~\cite{Libbrecht} and in the LASTI
prototype at MIT.

\section{Future work}\label{FutureWork}

The measurements and predictions described here indicate that mechanical loss associated
with dielectric optical coatings may be a significant source of thermal noise in advanced
LIGO. Plans are underway for experiments that will allow us to better understand and,
perhaps, reduce the coating thermal noise. A program of loss measurements on various
optical coatings deposited on both fused silica and sapphire substrates has begun so that
the most appropriate coating may be found.  There are also plans to try and correlate the
loss angle of the coating with other methods of interrogating its structure. To improve
the coating thermal noise without major changes to the optics, the coating loss must be
reduced.  Study of different dielectric materials is clearly warranted, and changes in
the deposition process or post-deposition annealing might also lead to improvements.  An
agreement has been reached between the LIGO laboratory and two optical coating companies
to engage in such research.

Two main models exist for understanding the source of the excess loss in the coating. One
is that the internal friction of the coating materials, thin layers of Ta$_2$O$_5$ and
SiO$_2$, is high.  The other model is that the excess damping comes from rubbing between
the layers, and between the coating and the substrate.  Experiments are underway to test
these models.

Measurement of the unknown parameters in \eqn{ThermalNoiseResult} are crucial. As
discussed in Sec.~\ref{Theory}, ringdown $Q$ measurements can not determine
$\phi_{\perp}$ due to the boundary conditions on the free vibration of a sample.  A
variation of the anelastic aftereffect experiment~\cite{AA}, which will measure the
relaxation rate of the coating after being stressed perpendicularly to the substrate, is
being pursued at Caltech~\cite{Willems}. This experiment should give a direct measurement
of $\phi_{\perp}$.

As seen in Eq.~(\ref{ThermalNoiseResult}), the coating thermal noise in an interferometer
is a strong function of the laser spot size.  Increasing the size of the laser spot
reduces the effect of the coating loss on the total thermal noise, so large spot sizes
are desirable.  Large spots also help decrease the effect of thermoelastic damping  in
sapphire mirrors~\cite{Braginsky}, so configurations to increase the spot size are
already being considered.  A spot size of about 6~cm is the largest that can be achieved
on the 25~cm diameter test masses while still keeping the power lost due to diffraction
below $\sim 15$~ppm. In the case of 25 cm diameter fused silica test masses, the largest
spot size that can be achieved is about 5.5~cm, limited by thermal
lensing~\cite{SystemsDesignDocument}. Larger diameter test masses and correspondingly
larger spot sizes would be one way to reduce the effects of the coating loss on advanced
LIGO's noise. However, this would require a re-evaluation of a number of advanced LIGO
subsystems.

\section*{Acknowledgments}
We would like to thank Helena Armandula and Jordan Camp for their help in getting our
samples coated.  We thank L.~Samuel Finn and everyone who contributed to {\tt BENCH}, and
for making it available to us.  Fred Raab first suggested that we look at the 40~m
prototype noise data, and we had useful discussions about the 40~m prototype with David
Shoemaker, Mike Zucker, Peter Fritschel, and Stan Whitcomb.  Peter Fritschel also showed
us how to optimize {\tt BENCH} for maximum binary neutron star reach and helped in
getting the large disk sample coated. Phil Willems provided thoughtful comments on the
limits of applicability of the coating loss model. We thank Eric Gustafson for useful
discussions about experiments with fused silica.  The Syracuse University glassblower,
John Chabot, gave crucial help by teaching us about glass welding.
We thank Mike Mortonson for help in the lab, and Emily~S.~Watkins for useful comments on
the manuscript. This work was supported by Syracuse University, U.S. National Science
Foundation Grant Nos. PHY-9900775 and PHY-9210038, the University of Glasgow, and PPARC.

\appendix

\section{Stresses and strains in the coating}\label{CoatingStressesAndStrainsAppendix}
\renewcommand{\arraystretch}{1}
We obtain the stresses and strains in the coating in terms of the stresses and strains in
the surface of the substrate by utilizing the thin coating approximation, and assuming
that the coating Poisson's ratio is not very different from that of the substrate.
Denoting strains by $\epsilon_{ij}$ and stresses by $\sigma_{ij}$, this can be summarized
in terms of the following constraints. In cylindrical coordinates,
\renewcommand{\arraystretch}{1.5}
\begin{equation}\label{Conditions}
\left.\begin{array}{lll}
&\epsilon_{rr}'          &=\epsilon_{rr}\\
&\epsilon_{\theta\theta}'    &=\epsilon_{\theta\theta}\\
&\epsilon_{rz}'          &=\epsilon_{rz}\\
&\sigma_{zz}'            &=\sigma_{zz}\\
&\sigma_{rz}'            &=\sigma_{rz}
\end{array}\right.
\end{equation}
where primed quantities refer to the coating and the unprimed quantities refer to the
surface of the substrate. Due to axial symmetry
$\epsilon_{r\theta}'=\epsilon_{z\theta}'=\sigma_{r\theta}'=\sigma_{z\theta}'=0$. We use
the following relations, valid for axially symmetric deformations~\cite{Bondu}
\begin{equation}\label{StressStrainRelations}
\left.\begin{array}{lll}
\sigma_{rr}&=&(\lambda+2\mu)\epsilon_{rr}+\lambda\epsilon_{\theta\theta}+\lambda\epsilon_{zz}\\
\sigma_{\theta\theta}&=&\lambda\epsilon_{rr}+(\lambda+2\mu)\epsilon_{\theta\theta}+\lambda\epsilon_{zz}\\
\sigma_{zz}&=&\lambda\epsilon_{rr}+\lambda\epsilon_{\theta\theta}+(\lambda+2\mu)\epsilon_{zz}\\
\sigma_{rz}&=&2\mu\epsilon_{rz}
\end{array}\right.
\end{equation}
where $\lambda$ and $\mu$ are the Lam\'{e} coefficients. In terms of  Young's modulus and
Poisson's ratio, the Lam\'{e} coefficients are
\begin{equation}\label{LameCoefficients}
\begin{array}{lll}
\lambda&=&Y\sigma/((1+\sigma)(1-2\sigma)),\\
\mu&=&Y/(2(1+\sigma)).
\end{array}
\end{equation}
Combining  Eqs.~(\ref{Conditions}) and Eqs.~(\ref{StressStrainRelations}), we obtain the
stresses and strains in the coating in terms of the stresses and strains in the surface
of the substrate \begin{equation}\label{CoatingInTermsOfSubstrate}
\left.\begin{array}{lll}
\epsilon_{rr}'          &=&      \epsilon_{rr}\\
\epsilon_{\theta\theta}'    &=&
\epsilon_{\theta\theta}\\
\epsilon_{zz}'&=&\frac{\lambda-\lambda'}{\lambda'+2\mu'}\,(\epsilon_{rr}+\epsilon_{\theta\theta})
+\frac{\lambda+2\mu}{\lambda'+2\mu'}\,\epsilon_{zz}\\
\epsilon_{rz}'&=&\epsilon_{rz}\\
\sigma_{rr}'&=&(\lambda'+2\mu')\epsilon_{rr}+\lambda'\epsilon_{\theta\theta}+\lambda'\epsilon_{zz}'\\
\sigma_{\theta\theta}'&=&\lambda'\epsilon_{rr}+(\lambda'+2\mu')\epsilon_{\theta\theta}+\lambda'\epsilon_{zz}'\\
\sigma_{zz}'            &=&      \sigma_{zz}\\
\sigma_{rz}'&=&\sigma_{rz}
\end{array}\right.
\end{equation}
where $\lambda'$ and $\mu'$ are the Lam\'{e} coefficients of the coating, and $\lambda$
and $\mu$ are the Lam\'{e} coefficients of the substrate.

\begin{figure}
\includegraphics[width=10cm]{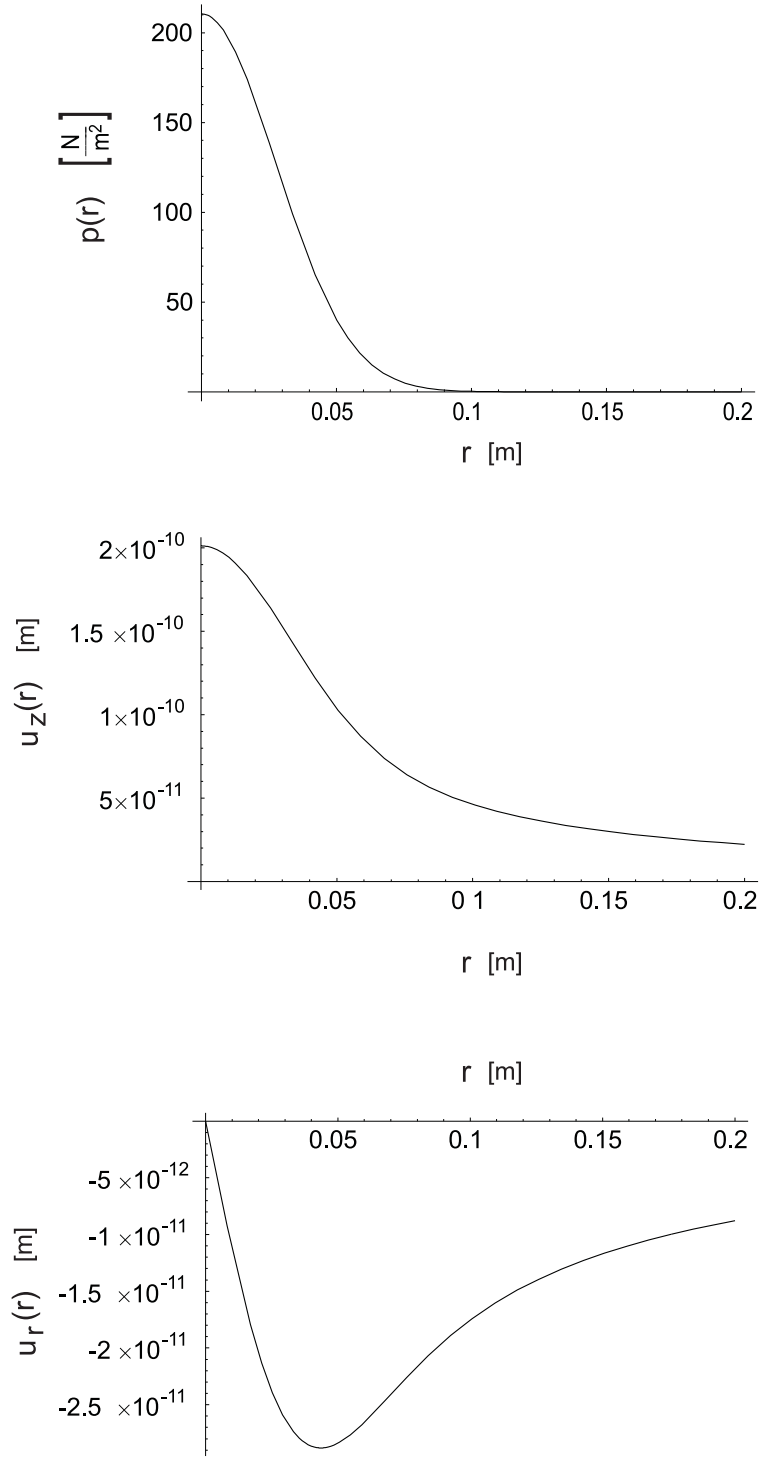}\vspace{2ex}
\caption{{\em Top:\/}~Pressure distribution $p(r)$ from
\eqn{PressureDistribution} (with $F$ set to unity). The pressure distribution has
the same shape as the laser intensity.~~{\em Center:\/}~The resulting response of the
surface in the axial direction, $u_z(r)$. The impression is wider than the applied
pressure.~~{\em Bottom:\/}~The resulting response in the radial direction, $u_r(r)$.  As
expected, the surface is pulled toward the center.}\label{Allfig}
\end{figure}

We obtain the stresses and strains in the substrate $\epsilon_{ij}$, $\sigma_{ij}$ from
the general solutions to the axially symmetric equations of elasticity for an infinite
half-space~\cite{Bondu,ThorneLiu}
\begin{equation}\label{AxissymElasticEqsSoln}
\begin{array}{lll}
u_r(r,z)&=&\int_0^\infty \,[\alpha(k)-\frac{\lambda+2\mu}{\lambda+\mu}\,\beta(k)+\beta(k)
k z]\,
e^{-kz}\!J_1(kr)\, k dk \\
u_z(r,z)&=&\int_0^\infty \,[\alpha(k)+\frac{\mu}{\lambda+\mu}\,\beta(k)+\beta(k) k z]\,
e^{-kz}\!J_0(kr)\, k dk \\
u_\theta(r,z)&=&0 \qquad \mbox{(Axial symmetry)},
\end{array}
\end{equation}
where $u_r(r,z)$ is the radial deformation of the test mass, $u_z(r,z)$ is the
deformation of the test mass perpendicularly to the face ($z$ being positive inward), and
$u_\theta(r,z)$ is the transverse displacement. $J_1(kz)$ and $J_0(kz)$ are Bessel
functions of the first kind. The functions $\alpha(k)$ and $\beta(k)$ are determined by
the boundary conditions at the front face: $\sigma_{rz}(r,z=0)=0$ and
$\sigma_{zz}(r,z=0)=p(r)$~\cite{Bondu}. Using the pressure distribution $p(r)$ from
Eq.~(\ref{PressureDistribution}) gives
\begin{equation}\label{alphak}
\alpha(k)=\beta(k)=\frac{F}{4\pi \mu \,k } \,\exp\left(-\frac{1}{8}\,k^2 w^2\right).
\end{equation}
Substituting \eqn{alphak} into Eqs.~(\ref{AxissymElasticEqsSoln}) and performing the
integrals leads to\vspace{1ex}
\begin{eqnarray}\label{rSurfaceDeformation}
u_r(r,z=0)&=&-\frac{F(\w)}{4\pi(\lambda+\mu)r}
\left[1-\exp\!\left(-\frac{2r^2}{w^2}\right)\right],
\\
u_z(r,z=0)&=&\frac{F(\w)\,(\lambda+2\mu)}{ 2\sqrt{2\pi}\,(\lambda+\mu)\mu w}
\;\exp\!\left(-\frac{r^2}{w^2}\right)I_0\!\left(\frac{r^2}{w^2}\right)
\label{zSurfaceDeformation}
\end{eqnarray}
\vbox{~~\newline\vspace{-2ex}}
where $I_0$ is a modified Bessel function of the first kind. These deformations are
shown, along with the pressure distribution $p(r)$, in
\fig{Allfig}. The strains in the substrate are obtained from the
relations
\begin{equation}\label{StrainsInTermsOfDeformations}
\begin{array}{lll}
\epsilon_{rr}&=&\delta u_r/\delta r\\
\epsilon_{\theta\theta}&=&u_r/r\\
\epsilon_{zz}&=&\delta u_z/\delta z\\
\epsilon_{rz}&=&(\delta u_z/\delta r + \delta
u_r/\delta z)/2.
\end{array}
\end{equation}
These strains can now be used to find the stresses in the surface of the substrate
through Eqs.~(\ref{StressStrainRelations}), and then to find the stresses and strains in
the coating through Eqs.~(\ref{CoatingInTermsOfSubstrate}).  The results for the surface
of the substrate are
\renewcommand{\arraystretch}{2}
\begin{equation}\label{TopmostSubstrateLayerStressesAndStrains}
\begin{array}{lll}
\epsilon_{rr}&=&
\frac{F}{4\pi(\lambda+\mu)}
\left(\frac{1}{r^2}
    \left(1-e^{-2r^2/w^2}
    \right)
    -\frac{4}{w^2}\,e^{-2r^2/w^2}
\right)
\\
\epsilon_{\theta\theta}&=&
-\frac{F}{4\pi(\lambda+\mu)}
\left(\frac{1}{r^2}
    \left(1-e^{-2r^2/w^2}
    \right)
\right)
\\
\epsilon_{zz}&=&
-\frac{F}{4\pi(\lambda+\mu)}
\left(\frac{4}{w^2}\,e^{-2r^2/w^2}
\right)
 \\
\epsilon_{rz}&=&0
\\
\sigma_{rr}&=&
\frac{F}{2 \pi(\lambda+\mu)}
\left(
   \frac{\mu}{r^2}
   \left(1-e^{-2r^2/w^2}
   \right)
   -\frac{4(\lambda+\mu)}{w^2}\,
   e^{-2r^2/w^2}
\right)
\\
\sigma_{\theta\theta}&=&
-\frac{F}{2 \pi(\lambda+\mu)}
\left(\frac{\mu}{r^2}
    \left(1-e^{-2r^2/w^2}
    \right)
    +\frac{4\lambda}{w^2}\, e^{-2r^2/w^2}
\right)\\
\sigma_{zz}&=&
-\frac{F}{2\pi}\,
\left(\frac{4}{w^2}\,e^{-2r^2/w^2}
\right)
\\
\sigma_{rz}&=&0
\\
\end{array}
\end{equation}
and for the coating
\begin{equation}\label{CoatingStressesAndStrains}
\begin{array}{lll}
\epsilon_{rr}'&=&\epsilon_{rr}
\\
\epsilon_{\theta\theta}'&=&\epsilon_{\theta\theta}
\\
\epsilon_{zz}'&=&
-\frac{F(2(\lambda+\mu)-\lambda')}{4\pi(\lambda+\mu)(\lambda'+2\mu')}
\left(\frac{4}{w^2}\,
    e^{-2r^2/w^2}
\right)
\\
\epsilon_{rz}'&=&0
\\
\sigma_{rr}'&=&
\frac{F}{2\pi(\lambda+\mu)(\lambda'+2\mu')}
\left(\frac{\mu'(\lambda'+2\mu')}{r^2}
    \left( 1-e^{-2r^2/w^2}
    \right)
    -\frac{4(\lambda'(\lambda+\mu)+2\mu'(\lambda'+\mu'))}{w^2}\,
    e^{-2r^2/w^2}
\right)
\\
\sigma_{\theta\theta}'&=&
-\frac{F}{2\pi(\lambda+\mu)(\lambda'+2\mu')}
\left(\frac{\mu'(\lambda'+2\mu')}{r^2}
    \left( 1-e^{-2r^2/w^2}
    \right)
    +\frac{4(\lambda'(\lambda+\mu+\mu'))}{w^2}\,
    e^{-2r^2/w^2}
\right)
\\
\sigma_{zz}'&=&\sigma_{zz}
\\
\sigma_{rz}'&=&0\,.
\end{array}
\end{equation}
Equations~(\ref{TopmostSubstrateLayerStressesAndStrains}) can now be used to find the
energy density in the substrate and integrated over the half-infinite volume,
\eqn{TotalEnergyDefn}, to give the total energy in the substrate, \eqn{TotalEnergy}.
Equations~(\ref{CoatingStressesAndStrains}) can be substituted into the expression for
the energy density at the surface, \eqn{EnergyComponents}, and integrated over the
surface to give the expressions for $\delta U_\|$ and $\delta U_\bot$ in
Eqs.~(\ref{dUpara})~and~(\ref{dUperp}).
\renewcommand{\arraystretch}{1}

\pagebreak[4]

\renewcommand{\arraystretch}{1.16}

\begin{table}
\begin{tabular}{ccccc}
\hline
{~Slide~} & {~Coating~} & {~Mode~} & {~Frequency~} & {~Q~}\\
\hline\hline
A & HR & 2 & 1022~Hz & $1.1 \pm 0.5 \times 10^5$ \\
  & HR & 3 & 1944~Hz & $1.6 \pm 0.1 \times 10^5$ \\
  & HR & 4 & 2815~Hz & $1.6 \pm 0.1 \times 10^5$ \\
B & HR & 2 &  962~Hz & $1.3 \pm 0.1 \times 10^5$ \\
C & none & 2 & 1188~Hz & $4.0 \pm 0.2 \times 10^6$ \\
  & none & 3 & 2271~Hz & $4.9 \pm 0.3 \times 10^6$ \\
\hline
\end{tabular}
\caption{Measured $Q$'s for transverse bending modes of the three commercially polished
fused silica slides.  Slides A and B were coated while slide C was left uncoated as a
control.}\label{table:slide}
\end{table}

\begin{table}
\begin{tabular}{cccc}
\hline
{~Hanging Number~} & {~Coating~} &
{~Frequency~} & {~Q~}\\
\hline\hline
1 & none & 4107~Hz & $3.46 \pm 0.02 \times 10^6$ \\
2 & none & 4107~Hz & $3.10 \pm 0.007 \times 10^6$ \\
3 & HR ($45^{\circ}$) &  4108~Hz & $1.28 \pm 0.02 \times 10^6$ \\
4$^\dagger$ & HR ($45^{\circ}$) &  4121~Hz & $1.24 \pm 0.001 \times 10^6$\\
\hline
\multicolumn{4}{l}{\raisebox{0pt}{\footnotesize $^\dagger$Ear was sheared off twice before this
hanging.}}
\end{tabular}
\caption{Measured $Q$'s for butterfly mode of the superpolished fused silica disk.
In hangings~1~and~2, the disk remained uncoated whereas in hangings~3~and~4 the disk had
been coated.}\label{table:thickdisk}
\end{table}

\begin{table}
\begin{tabular}{cllll}
\hline
{~Sample~} & {~Parameter~~~~~~~~~~~~~~~~~~~~~~~~~~~~~~}&
{Value~~~~~~~~~~~}&{Units~} \\
\hline\hline
Slide & Coating Layers        & 14 &\\
      & Coating Thickness $d$ &  2.4~&$\!\mu$m \\
Disk  & Coating Layers        & 38 &\\
      & Coating Thickness $d$ &  24.36~&$\!\mu$m \\
Both  & Substrate Young's modulus ($Y$)& $7.0 \times 10^{10}$~&N/m$^2$ \\
      & Coating Young's modulus ($Y'$) & $7 \times 10^{10}$~&N/m$^2$\\
\hline
\end{tabular}
\caption{
Physical parameters of the coating and samples. These values are used to calculate the
coating loss $\phi_\|$ from \eqn{RingdownPhi}.  The value of the coating Young's modulus
is a preliminary result from SMA/Virgo in Lyon, France.}\label{table:params}
\end{table}

\begin{table}
\begin{tabular}{llrc}
\hline
Test mass~~~~~~ &                      &~~~~$Q\sub{eff}~~~~~~$                   &~~~~~~Structural thermal noise\\
material~~~~~~~ &  Coating loss~~~~~~~ &~(={$1/\phi_{\mathrm{readout}}$})~&~~~~~~at 100 Hz,~~$\sqrt{S_h}$\\
\hline
\hline
Sapphire     &none                      & $200 \times 10^6$  &~~~~~$1 \times 10^{-24}$  \\
             &$\phi_\|=1\times 10^{-4}$ & $15 \times 10^6$   &~~~~~$3 \times 10^{-24}$ \\
             &$\phi_\|=4\times 10^{-4}$ & $4 \times 10^6$    &~~~~~$5 \times 10^{-24}$\\
\hline
Fused silica &none                      & $30 \times 10^6$   &~~~~~$6 \times 10^{-24}$ \\
             &$\phi_\|=1\times 10^{-4}$ & $19 \times 10^6$   &~~~~~$7 \times 10^{-24}$\\
             &$\phi_\|=4\times 10^{-4}$ & $9 \times 10^6$    &~~~~~$9 \times 10^{-24}$ \\
\hline
\end{tabular}
\caption{Comparison of structural thermal noise with and without taking coatings into
account.  The effective quality factor $Q\sub{eff}$ (equal to the reciprocal of
$\phi_{\mathrm{readout}}$) represents the quality factor a homogeneous mirror would need
to have to give the same structural contribution to thermal noise as the actual coated
mirror. (The effect of thermoelastic damping, important for sapphire, is not included in
$Q\sub{eff}$). The final column shows the strain amplitude thermal noise at 100 Hz in the
advanced LIGO interferometer resulting from structural loss in the test mass coatings and
substrates.}\label{table:bench}
\end{table}

\renewcommand{\arraystretch}{1}

\end{document}